\documentclass{PoS}

\newcommand{\eq}{\begin{equation}} 
\newcommand{\en}{\end{equation}} 

\def\one{{\rm 1\kern -.9mm l}}                             %

\newcommand{\eqa}{\begin{eqnarray}}
\newcommand{\ena}{\end{eqnarray}}
\def\beq{\begin{equation}}
\def\eeq{\end{equation}}
\def\beq{\begin{equation}}
\def\eeq{\end{equation}}
\def\beqa{\begin{eqnarray}}
\def\eeqa{\end{eqnarray}}

\def\ii{\mathrm{i}}
                                                           
\newcommand{\p}{\partial}

\def\ii{\mathrm{i}}

\newcommand{\cE}{\mathcal{E}}

\newcommand\NN{{\mathbb N}}

\newcommand{\bL}{\bar{L}}
\newcommand{\bep}{\bar{\epsilon}}
\newcommand{\ep}{\epsilon}

\newcommand{\bE}{\bar{\cal E}}

\newcommand{\bn}{\bar{n}}

\newcommand{\ga}{{\cal C}}
\newcommand{\bga}{{\bar{\cal C}}}

\newcommand{\Log}{{ln}}

\newcommand{\cIR}{{c}_{{IR}}}
\newcommand{\tcIR}{\tilde{{c}}_{{IR}}}

\title{Recent progress in the effective string theory description of LGTs.}

\ShortTitle{Recent progress in the effective string theory description of LGTs.}

\author{{Marco Bill\`o$^a$},~~ \speaker{Michele Caselle}$^a$,~~ Davide Fioravanti$^c$,~~Ferdinando Gliozzi$^{a,b}$,~~Marco Meineri$^d$,
~~Roberto Pellegrini$^a$,~~ Roberto Tateo$^a$\\
        $^a$ Dipartimento di Fisica, Universit\`a di Torino \\
        and Istituto Nazionale di Fisica Nucleare, sezione di Torino\\
        via P. Giuria 1, 10125 Torino (Italy)\\
	$^b$ School of Computing and Mathematics \& Centre for Mathematical Science,\\
Plymouth University, Plymouth PL4 8AA, UK\\
$^c$ Istituto Nazionale di Fisica Nucleare - sezione di Bologna  \\
and Dipartimento di Fisica e Astronomia, Universit\`a di Bologna,\\
Via  Irnerio 46, I- 40127 Bologna, Italy\\
$^d$ Scuola Normale Superiore, Piazza dei Cavalieri 
7 I-56126 Pisa, Italy\\
and Istituto Nazionale di Fisica Nucleare - sezione di Pisa\\
        E-mail: \email{(billo)(caselle)(gliozzi)(ropelleg)(tateo)@to.infn.it}\\
        \email{fioravanti@bo.infn.it}, ~~ \email{marco.meineri@sns.it}}

\abstract{In presence of a static pair of sources, the spectrum of low-lying states
of any confining gauge theory in D space-time dimensions is
described, at large source separations, by an effective string theory.
Recently two  important advances improved our understanding of this
effective theory.
First, it was realized that the form of the effective action is strongly  
constrained by the requirement of the Lorentz invariance of the gauge theory,
 which is spontaneously broken by the formation
of a long confining flux tube in the vacuum. This constraint is strong
enough to fix uniquely the first few subleading terms of the action.      
Second, it has been realized that the first of these allowed terms - a
quartic polynomial in the field derivatives - is exactly
the composite field $T\bar{T}$, built with the chiral components, $T$ and
$\bar{T}$, of the energy-momentum tensor of the 2d QFT describing 
the infrared limit of the effective string.
This irrelevant perturbation is quantum integrable and yields, through the
thermodynamic Bethe Ansatz (TBA), the energy
levels of the string which exactly coincide with the Nambu-Goto spectrum. 
In this talk we first review the general implications of these two results
and then, as a test of the power of these methods,
use them to construct the first few boundary corrections to the effective
string action.}

\FullConference{31st International Symposium on Lattice Field Theory - LATTICE 2013\\
		July 29 - August 3, 2013\\
		Mainz, Germany}

\begin{document}
\section{Introduction}
It is widely believed that the dynamics of the flux tube joining a quark-antiquark pair is 
well described by an effective string action
$S$ which  
flows at large scales towards a  
massless free-field theory. Thus, for large enough inter-quark separations 
it is not necessary to know explicitly the specific form of the effective string action $S$, but only its infrared limit 
\cite{Luscher:1980fr}
\beq
S[X]=S_{cl}+S_0[X]+\dots,
\label{frees}
\eeq
where the classical action $S_{cl}$ describes the usual perimeter-area term,
$X$ denotes the 2d bosonic 
fields $X_i(\xi_1,\xi_2)$, with  $i=1,2,\dots, D-2$,     
describing the 
transverse displacements of the string with respect the configuration 
of minimal energy, $\xi_1,\xi_2$ are the coordinates on the world-sheet
and $S_0[X]$ is the Gaussian action
\beq
S_0[X]=\frac\sigma2\int d^2\xi\left(\partial_\alpha X\cdot\partial^\alpha X
\right) ~.
\label{gauss}
\eeq
In the past few years there has been  substantial progress in our understanding
of the effective string action both from the theoretical and from the numerical point of view. 
In particular we want to discuss in this short review  two major advances:
   
   \begin{itemize}
    \item  The Effective String action is strongly constrained by Lorentz invariance. 
    The first few corrections in eq. (\ref{frees}) are uniquely fixed and coincide with those 
    arising from the Nambu-Goto action. This explains why Nambu-Goto
    describes so well the infrared regime of Wilson loops or Polyakov loop correlators.

    \vspace{2mm}
    \item The Nambu-Goto effective string can be described as a free 2d bosonic theory  perturbed by the irrelevant operator $T\bar T$
    (where $T$ and $\bar T$ are the two chiral components of the energy momentum tensor). This perturbation is quantum
    integrable and yields, using the Thermodynamic Bethe Ansatz (TBA), a spectrum which, in a suitable limit, coincides with the
    Nambu-Goto one. 
  \end{itemize}

As an example of the possible applications of these results in the final part of this paper we shall use them to
study  the boundary corrections to the effective string action and evaluate their contribution to Wilson loops and 
Polyakov loop correlators. We shall then test these predictions with high precision numerical estimates in the 3D gauge Ising model.

\section{Lorentz invariance}
The Effective String action may contain only
terms respecting the internal and space-time symmetries of the system.
It can be written as a low energy expansion in the number of derivatives of the $X$ fields. In particular the 
first few terms beyond the free 2d bosonic theory of eq. (\ref{gauss}) for an open string stretched between fixed ends, 
for instance Polyakov loops, are \cite{Luscher:2004ib}
 \beq
S=S_{cl}+\frac\sigma2\int d^2\xi\left[\partial_\alpha X\cdot\partial^\alpha X+
c_2(\partial_\alpha X \cdot\partial^\alpha X)^2
+c_3(\partial_\alpha X \cdot\partial_\beta X)^2+\dots\right]+S_b\,,
\label{action}
\eeq
where $S_b$ is the boundary contribution characterizing the open string. If the boundary is a Polyakov line in the $\xi_0$ 
direction placed at $\xi_1=0$, on which we assume  Dirichlet boundary conditions
$X_i(\xi_0,0)=0$, the first few terms contributing to  $S_b$ are
\beq
S_b=\int d\xi_0 \left[b_1\partial_1 X\cdot \partial_1 X+b_2\partial_1\partial_0 X\cdot \partial_1\partial_0 X+b_3(\partial_1 X\cdot \partial_1 X)^2+
\dots\right]\,.
\label{bounda}
\eeq
As first observed in 2004 by L\"uscher and Weisz \cite{Luscher:2004ib} the coefficients $c_i$ and $b_i$ of the above expansion 
should satisfy some consistency constraints; they were obtained by the comparison of the string partition function 
in different channels (``open-closed string duality''). These results were further generalized in \cite{Aharony:2009gg}. 
However it was later realized 
\cite{{Meyer:2006qx},{Aharony:2010cx}} that the crucial ingredient of these 
constraints is the Lorentz symmetry of the underlying Yang-Mills theory.
Indeed, even if the complete $SO(1,D-1)$  
invariance is broken by the classical configuration around which we 
expand, the effective action should still respect this symmetry through a 
non-linear realisation in terms of the transverse fields $X_i$. In this way 
it was shown \cite{{Aharony:2010cx},{Gliozzi:2011hj}} that the terms with
 only first derivatives  coincide with the Nambu-Goto action to all orders in the derivative expansion.
  The first allowed  correction to the Nambu-Goto 
action  in the bulk turns out to be the the six derivative term  \cite{Aharony:2009gg}
\beq
c_4\left(\partial_\alpha\partial_\beta X\cdot\partial^\alpha\partial^\beta X
\right)\left(\partial_\gamma X\cdot\partial^\gamma X\right)
\eeq
with arbitrary coefficient $c_4$; however this term is non-trivial only 
for $D>3$. At $D=3$ the first non-trivial deviation to the Nambu-Goto action 
is an eight-derivative term and it has been recently shown \cite{Aharony:2011gb}, 
using the recursion relations generated by the non-linear Lorentz 
transformations, that it generates a geometric term proportional to the 
squared curvature of the induced metric on the world-sheet. These recursion relations and their 
relationship with diffeomorphism invariance were further investigated in \cite{Gliozzi:2012cx, Cooper}, 
while the link
with the orthogonal gauge (Polchinski-Strominger) approach to the effective action was studied 
in \cite{Dubovsky:2012sh} (see also \cite{Aharony:2013ipa}).
The fact that the first deviations from the Nambu-Goto string are of 
high order, especially in $D=3$, explains why in early Monte Carlo 
calculations \cite{{Caselle:1994df},{Caselle:2005xy},{Caselle:2006dv}}  a good agreement 
with the Nambu-Goto string was observed.
A similar analysis can be performed for the the effective string spectrum 
\cite{Aharony:2010db}. Also in this case, for the majority of the states, 
a good agreement with lattice simulations was found 
\cite{{Athenodorou:2011rx},{Athenodorou:2010cs}}. 

\section{The Nambu-Goto action as the $T\bar T$ perturbation of $S_0$.}

The energy-momentum tensor of the free-field theory (\ref{gauss}) can be written as
\beq
T_{\alpha\beta}=\p_\alpha X\cdot\p_\beta X-\frac12\delta_{\alpha\beta}
\left(\p^\gamma X\cdot\p_\gamma X\right)\,.
\label{Tab}
\eeq
Setting in eq. (\ref{action}) the values $c_2=\frac18$ and $c_3=-\frac14$ prescribed by Lorentz invariance, we find
\beq
S=S_{cl}+S_0[X]-\frac\sigma4\int d^2\xi\, T_{\alpha\beta}T^{\alpha \beta}+S_{b}+\dots
\label{sat}
\eeq
In 2d CFTs it is useful to introduce the chiral components 
$T_{zz}=\frac12(T_{11}-iT_{12})$ and $T_{\bar{z}\bar{z}}=
\frac12(T_{11}+iT_{12})$ and use the normalized quantities 
$T=-2\pi\sigma T_{zz}$, $\bar{T}=-2\pi\sigma T_{\bar{z}\bar{z}}$ in such a way the operator product expansion begins with
\beq
T(z)T(w)=\frac{D-2}2\frac1{(z-w)^4}+\dots
\eeq 
and similarly for $\bar{T}$.
Thus we may write
\beq
S=S_{cl}+S_0[X]-\frac1{2\pi^2\sigma}\int d^2\xi\, T \bar{T}+S_{b}+\dots
\label{SN}
\eeq
This shows that the (universal) effective string theory at the next to leading order is nothing else than the $T\bar T$ perturbation of
the free field theory \cite{Caselle:2013dra}.  This massless perturbation can be studied using TBA techniques. 
For the ground state with periodic boundary conditions, this  was pioneered  
several years ago by Zamolodchikov
\cite{Zamolodchikov:1991vx}. 
Let us briefly review  the TBA analysis of \cite{Zamolodchikov:1991vx, Dubovsky:2012wk, Caselle:2013dra}. 
The simplest instance discussed in  
\cite{Zamolodchikov:1991vx} corresponds 
to the RG  flow  connecting  the tricritical Ising to the  Ising model and it is described 
by a  single species  of  interacting  massless Majorana fermions. Close to  the IR fixed point the model
is described by   an effective action of the form (\ref{SN}) with $S_{b}=0$:
\beq
S= S_{Ising}-\frac1{2\pi^2\sigma}\int d^2\xi\, T \bar{T}+\dots~,~~(\sigma \gg 1)~,
\label{ising}
\eeq
where $\sigma$ sets the  scale of the perturbation and will play the role of string tension 
in our effective string interpretation. In 1+1-dimensions, massless excitations confined on a ring  
naturally separate  into right and left movers.  In the case under consideration, the right-right
and left-left mover scattering  is trivial  while the left-right  scattering is described by the  amplitude
\beq
S(p,q)=\frac{2 \sigma+ i p q }{2 \sigma- i  p q }~,
\label{sm}
\eeq
where $p$ is the momentum of the right mover and $-q$ the
momentum of the left mover.
Starting from the   $S$-matrix (\ref{sm}),  Zamolodchikov was able to derive the
TBA  equations for the vacuum energy of the theory defined on a 
infinite cylinder  with circumference $R$. Considering also  the excited states 
\cite{Bazhanov:1996aq,Dorey:1996re},  
the TBA  equations are 
\beq
{\ep(p)= R p - \int_{{\bar{\cal C}}}  \frac{dq}{2 \pi} \phi(p,q)\,\bL(q)
,~~~
\bep(p)= R p - \int_{{\cal C}}\frac{dq}{2 \pi} \phi(p,q)\,L(q)~,}
\label{tba0}
\eeq
where $\ep(p)$ and $\bep(p)$ are the   pseudoenergies for the right and
the left movers, respectively, and
\beq
\phi(p,q)= -i \partial_q \Log  S(p,q)~,~~ L(p)=\Log(1+ e^{-\ep(p)})~,~~\bL(p)= \Log(1+ e^{-\bep(p)})~.
\nonumber
\eeq
In (\ref{tba0}), the integration contours $\ga$ and $\bga$ run  from $q=0$ to $q=\infty$  on the real axis 
for the ground state, but for excited states  they circle around a finite number of poles 
$\{ q_i \}$ and $\{ \bar{q}_i \}$  
 of $\partial_q L(q)$ and $\partial_q \bL(q)$ (cf. \cite{Dubovsky:2012wk, Caselle:2013dra}). The
corresponding energy is 
\begin{equation}
E^{(TBA)}(R,\sigma)= - \int_{{\cal C}}\frac{dp}{2\pi}  L(p) -  \int_{{\bar{\cal C}}} \frac{dp}{2\pi} \bar{L}(p)~. 
\label{enn}
\end{equation}
{} From (\ref{tba0}-\ref{enn}), the  $T\bar T$ contribution to the energies   can be spotted only perturbatively 
\cite{Zamolodchikov:1991vx}, but if we consider  only  
the leading part $S_1(p,q)$
of the  Zamolodchikov's $S$ matrix  at large  $\sigma$
\beq
{S(p,q)= e^{i pq/\sigma -i (pq/\sigma)^3/12 + \dots   } = S_1(p,q) e^{-i (pq/\sigma)^3/12 + \dots}}
\eeq
then the kernel becomes
$ \phi(p,q)=-i \partial_q \Log\; S_1(p,q) =  p/\sigma$,
and the corresponding  TBA equations can be solved explicitly \cite{Dubovsky:2012wk}.  In fact,  
it is easy to show  that they are fully equivalent to 
the following simple algebraic equations \cite{Caselle:2013dra}:
\beq
E_{(n,\bn)}(R, \sigma)= E^{(TBA)}(R,\sigma)+ \sigma R = \cE +\bE+\sigma R~,~
\eeq
{\beq
\cE= - \frac{\pi (\tcIR-24n)  }{12 (R + \bE/\sigma)}~,~~~
\bE= - \frac{\pi (\tcIR-24\bar{n})}{12 (R+\cE/\sigma)}~,
\eeq}
with $n,\bn \in \NN$.
These equations  can be  easily solved, giving
{\beq
E_{(n,\bn)}(R,\sigma) =  \pm   \sqrt{ \sigma^2 R^2 +
4 \pi \sigma \left( n +\bn - \frac{\tcIR}{12} \right)
+\left( \frac{ 2 \pi ( n -\bar{n})}{R} \right)^2
} \,\, .
\eeq}
Setting {$\tcIR=D-2$} this becomes exactly the N-G spectrum. For the massless flow (\ref{ising}) 
$\tcIR=c_{Ising}=1/2$, 
while for the single boson field theory discussed in \cite{Dubovsky:2012wk}  $\tcIR=1$. However,
it is interesting to notice that
the result  is more general and holds not only for free field theories but also
for more general  infrared CFTs \cite{Caselle:2013dra}.
The only change is that $\tcIR=\cIR-24h$ where $h$ may be anyone 
of the conformal weights of the theory (and can be tuned using suitable boundary conditions).

\section{Example: the boundary term}
Using the results of the previous sections we can fix the expression of the first few terms of the
boundary action eq. (\ref{bounda}) and find its contribution to the effective string spectrum. 
It was already realized in \cite{Luscher:2004ib} that $b_1=0$. It was later observed in \cite{Aharony:2010cx} 
that also $b_3=0$ and only the $b_2$ term is compatible with Lorentz invariance. In \cite{Billo:2012da} we
completed the analysis by solving the recursion relations dictated by the requirement of Lorentz invariance. 
The resulting expansion can be written in the  closed form
\beq
b_2\int d\xi_0 \left[
\frac{\partial_0\partial_1 X\cdot\partial_0\partial_1 X}{1+\partial_1 X\cdot\partial_1X}-
\frac{\left(\partial_0\partial_1 X\cdot\partial_1 X\right)^2}
{\left(1+\partial_1 X\cdot\partial_1X\right)^2}\right]\,.
\label{firstb}
\eeq 
In order to estimate the coefficient $b_2$ we performed a set of high precision simulations in the 3D gauge Ising model, evaluating both Wilson loops and Polyakov loop
correlators. Within the precision of our data we could measure only the contribution of the first term in the expansion 
of eq. (\ref{firstb}):
\begin{equation}
\label{derexpsb1}
S_{b,2}^{(1)} = b_2 \int_{\partial \Sigma} (\partial_0\partial_1 X)^2~.
\end{equation}
The contribution of this term to the interquark potential (the ground state of the effective string spectrum)
depends on the geometry of the observable.
In the case of  Polyakov loop correlators we have \cite{Aharony:2010cx}
\beq
\label{polybound}
\langle S^{(1)}_{b,2} \rangle_P=-b_2\frac{\pi^3 L}{60 R^4} E_4(\ii\frac{L}{2 R})~,
\eeq
 where $R$ is the distance between
the two Polyakov loops and $L$ the length of the compactified time direction (i.e. the inverse temperature). 

For the Wilson loop we have \cite{Billo:2012da}
\begin{equation}
\label{wilsbound}
\langle S^{(1)}_{b,2} \rangle_W=-b_2 \frac{\pi^3}{60}\left[\frac{R}{L^4}
E_4\left(\ii\frac{R}{L}\right) + \frac{L}{R^4} E_4\left(\ii\frac{L}{R}\right)\right]~,
\end{equation}
where the Eisenstein function $E_4$ is defined as:
\begin{equation}
\label{defeis} 
E_{4}(\tau) = 1 + 240 \sum_{n=1}^\infty \sigma_{3}(n)
q^{n}~,
\end{equation}
where $q=\exp(2\pi\ii\tau)$ and $\sigma_p(n)$ is the sum of the $p$-th powers of the divisors of $n$:
\begin{equation}
\label{defsigma}
\sigma_p(n) = \sum_{m|n} m^p~. 
\end{equation}

In both cases these expressions turn out to fit remarkably well the numerical data allowing to obtain a rather precise estimate for $b_2$. The results for the three
different values of $\beta$ that we simulated in \cite{Billo:2012da} are reported in table \ref{bres}. 

\begin{table}[ht]
\centering
\begin{tabular}{|cccc|}
\hline
data set & $ b_2 $ & $ b_2 \sqrt{\sigma}^3 $  & $ \chi^2 $    \\
\hline
 1  & 7.25(15) & 0.0250(5) & 1.2 \\
 2  & 26.8(8)  & 0.0289(9) & 1.8 \\
 3  & 57.9(12) & 0.0319(7) & 1.3 \\
\hline
\end{tabular}
\caption{Values of $b_2$ obtained from the fits.}
\label{bres}
\end{table}

Looking at the third column we see that the data show a rather good scaling behaviour and allow us to propose as a final estimate for the $b_2$ parameter in the 3D
gauge
Ising model the value: $ b_2 \sqrt{\sigma}^3 \sim 0.032(2)$. 
 It is interesting to notice
 that this result is of the same order of magnitude (but opposite in sign) of the value of $b_2$ measured in the 3D $ SU(2) $ 
 gauge model in \cite{Brandt:2010bw}. 

\section{Acknowledgements}

M. M. was a visiting graduate fellow at the Perimeter Institute for Theoretical Physics (Waterloo, Ontario) while this work was completed. Research at Perimeter Institute is supported by the Government of Canada through Industry Canada and by the Province of Ontario through the Ministry of Economic Development and Innovation.
This work is supported in part by the Compagnia di San Paolo contract
"Modern Application in String Theory" (MAST) TO-Call3-2012-0088.

\end{document}